\documentclass[
 aps,
 11pt,
 final,
 notitlepage,
 oneside,
 twocolumn,
 nobibnotes,
 nofootinbib,
 superscriptaddress,
 noshowpacs,
 centertags]
{revtex4}
\usepackage{graphicx}
\usepackage[utf8]{inputenc}
\begin{document}
%\selectlanguage{english}

\title{Globular clusters as indicators of Galactic evolution}
\author{\firstname{N.~R.}~\surname{Arakelyan}
  \footnote{E-mail: n.rubenovna@mail.ru}}
\author{\firstname{S.~V.}~\surname{Pilipenko}}
\affiliation{Lebedev Physical Institute, Russian Academy of Sciences, Moscow, 117997  Russia}

\label{firstpage}

\begin{abstract}
We have studied the system of globular clusters (GCs) that formed in other galaxies and eventually accreted onto the Milky Way. Thus, the samples of GCs belonging to different tidal streams, obtained on the basis of the latest data from the Gaia observatory, were taken from the literature. We measured the anisotropy of the distribution of these GCs using the gyration tensor and found that the distribution of GCs in the streams is isotropic. Nevertheless, it can be seen that some of the accreted GCs included into existing samples actually belong to the disk of the Galaxy. To clarify the origin of GCs, we investigated the ``age--metallicity'' relation. This dependence demonstrates bimodality and its two different branches clearly show the difference between the clusters formed in the streams and in the disk of the Galaxy. Furthermore, we have studied the influence of the large--scale environment of the Galaxy (i.e., the Local Supercluster) on the distribution of satellite galaxies and Galactic GCs. The satellite galaxies of the Milky Way are known to form an anisotropic planar structure, so we included them in our analysis too. An inspection has shown that the plane of the satellite galaxies is perpendicular both to the disk of the Galaxy and the supergalactic plane. For GCs more distant than 100~Kpc, a similar picture is observed.

\vspace{10pt}
\end{abstract}

\maketitle
\section{Introduction} 
Globular   clusters   (GCs)   belong   to   the   oldest objects  inhabiting  galaxies.  According  to  the  hierarchical  theory,  the  galaxies  are  formed  by  merger  of low--mass  and  then  by  larger  mass  objects  \citep{1974ApJ...189L..51P}.  Whenthe galaxies of very different masses merge, as a rule, the lower mass galaxy disrupts gradually. As it continues its orbital motion, a tidal tail of gas, dust, stars, and GCs forms behind the galaxy, thus enriching the larger mass  galaxy.  This  type  of  mergers  have  occurred  and still  occur  with  our  Galaxy  too.  According  to  Bland--Hawthorn and Gerhard \citep{2016ARA&A..54..529B}, about 100 satellite galaxies have accreted onto the Galaxy during the life time of the Universe. But out of these satellite galaxies, only massive ones contribute GCs to our Galaxy, since galaxies  with  stellar  mass  $10^7$ $M_\odot$  have  very  few  GCs \citep{2018MNRAS.481.5592F}. Myeong et al. \citep{2018ApJ...863L..28M} argue that the clusters with the critical energy $E\geq-1.6\times10^5$   km$^2$ s$^{-2}$   were accreted from dwarf galaxies. Cosmological hydrodynamic  simulations  show  that  15--40\%  of  the  stars  in the  halo  were  formed  outside  the  Galaxy  (ex--situ), that  is,  in  dwarf  satellite  galaxies,  and  then  were accreted  \citep{2018MNRAS.473.4077P,2021arXiv210112216R}.  In  the  results  of  studies  by  different authors, the percentage of GCs formed ex--situ differs. For  example,  Forbes  \citep{2020MNRAS.493..847F}  claims  that  54\%  of  GCs (87 out of 160 clusters) were formed ex--situ and then accreted, while Kruijssenn et al. \citep{2019MNRAS.486.3180K} claim that the percentage of accreted clusters is 43. According to Massari et al. \citep{2019A&A...630L...4M}, this value reaches 60\%.

Thus,  a  significant  part  of  the  GCs  of  the  Galaxy was accreted from the outside. Information about the origin of GCs can be preserved both in the properties of the stellar population of GCs and in the spatial distribution and dynamics of GCs themselves. In particular,  it  is  well  known  that  both  satellite  galaxies  and GCs exhibit a disk--like structure perpendicular to the disk  of  the  Galaxy  \citep[see, e.g.][]{2005A&A...431..517K,2008ApJ...680..287M,2018MNRAS.481..918A}.  This structure  may  be  a  result  of  the  accretion  of  several galaxies that arrived in our Galaxy mainly from polar directions.  According  to  Zeldovich  theory  \citep{1970A&A.....5...84Z},  formation  of  the  large--scale  structure  of  the  Universe occurs through independent contraction or expansion of  the  matter  in  the  three  mutually  perpendicular directions. A good example of a structure contractingin one direction and expanding in two other directions is the Local Supercluster of Galaxies, which looks likea typical Zeldovich ``pancake''. This structure sets the preferred  direction  in  the  vicinity  of  the  Galaxy,  and therefore   can   affect   the   preferrential   direction   of accretion and distribution of accreted material in our Galaxy.

Tidal streams of the Galaxy are actively discussed in the literature \citep[e.g.][]{1994Natur.370..194I,1996ApJ...459L..73M,1999Natur.402...53H,2002ApJ...569..245N,2004AJ....128..245M,2004ApJ...615..732R,2006ApJ...642L.137B,2006ApJ...645L..37G,2006ApJ...641L..37G,2006ApJ...636L..97D, 2007ApJ...667L..57S,2007ApJ...658..337B,2008MNRAS.389.1391S,2009ApJ...698..567S,2009ApJ...700L..61N,2009AJ....137.3809C,2010ApJ...718.1128L,2010ApJ...711...32N, 2011ApJ...728..102W,2014MNRAS.445.2971C,2017A&A...600A.118C,2017MNRAS.467.1329N}. Recent measurement of the GCs proper motions using GAIA data made it possible to identify GCs belonging to specific tidal streams. The problem regarding the difference in the physical properties of GCs formed in--situ and ex--situ was studied in detail in \citep{2020ARep...64..805M}. It was shown there that accreted GCs differ by the abundance of alpha--elements, as well as by the range of masses. The purpose of our study is to check the spatial orientation of the GCs system belonging to the streams, i.e., admittedly accreted onto the Galaxy from outside. For this, the orientation of the systems identified by different authors was compared with the disk of the Galaxy, as well as with the plane of the Local Supercluster. In addition, the ``age–metallicity'' relation (AMR) for GCs belonging to the streams was analyzed and the relation between GCs colors and their origin was discussed.

The paper is organized as follows. In Section~2, the studied GC samples are described and the anisotropy of their distribution is investigated. In Section~3 the AMR for GCs is discussed. In Section~4, the influence of the Local Supercluster is measured. Conclusions are presented in Section~5.

\section{MILKY WAY GLOBULAR CLUSTERS IN TIDAL STREAMS}\label{sec:2}
Our Galaxy contains at least 157 GCs \citep[][2010 edition]{1996AJ....112.1487H}\footnote{\url{http://physwww.mcmaster.ca/~harris/Databases.html}} and \citep{2013ApJ...772...82H}. As time goes by, the papers regarding new clusters in the Milky Way appear (e.g., $FSR$~1716 \citep{2017ApJ...838L..14M}, $FSR$~1758 \citep{2019MNRAS.488.1235M,2019ApJ...870L..24B}, $VVV-CL001$ \citep{2011A&A...527A..81M}, $VVV-CL002$ \citep{2011A&A...535A..33M}, $BH$~140 \citep{2018A&A...618A..93C}, $Gran$~1 \citep{2019A&A...628A..45G}, $Pfleiderer$~2 \citep{2009AJ....138..889O}, $ESO$~93--8 \citep{1999A&AS..136..363B}, $Mercer$~5 \citep{2005ApJ...635..560M}, $Segue$~3 \citep{2010ApJ...712L.103B}, $Ryu$~059, $Ryu$~879 \citep{2018ApJ...863L..38R}, $Kim$~3 \citep{2016ApJ...820..119K}, $Crater$/$Laevens$~1 \citep{2014MNRAS.441.2124B,2014ApJ...786L...3L}, $Laevens$~3 \citep{2015ApJ...813...44L} and $BLISS$~1 \citep{2019ApJ...875..154M}). Although even before obtaining high--precision GAIA data, there were attempts to identify GCs belonging to the tidal streams \citep{2004MNRAS.348...12M,2004AJ....127.3394F,2010MNRAS.404.1203F,2014MNRAS.445.2971C}, but after the appearance of GAIA data, these attempts significantly advanced \citep{2018ApJ...863L..28M,2019MNRAS.488.1235M,2019A&A...630L...4M,2019ARep...63..274M,2019AstBu..74..403M,2020MNRAS.493..847F,2020AstBu..75..394A,2021MNRAS.508L..26P}. In this paper, we draw our attention to three studies: \citep{2019A&A...630L...4M} (hereinafter, Massari), \citep{2019MNRAS.488.1235M} (hereinafter, Myeong), and \citep{2020MNRAS.493..847F} (hereinafter, Forbes), which contain the most complete lists of GCs belonging to different tidal streams.

Forbes et al. list 76 clusters that belong to five progenitors--satellite galaxies. To check membership, the authors used integrals of motion (IOM), AMR, and alpha--elements dependencies. Out of these 76 GCs, nine belong to the well--known Sagittarius dwarf spheroidal galaxy (Sgr dSph), 28 belong to the Gaia--Enceladus dwarf galaxy, nine GCs -- to the Sequoia dwarf galaxy, 21 clusters -- to the low--energy satellite Koala, nine clusters -- to a low--mass satellite, the Helmi streams. Out of the remaining 36 clusters, they are either ``Low--energy'' ones (25 objects) or ``High--energy'' ones (11 clusters). These clusters have high energies and a wide range of angular momenta, which suggests that they originated from different progenitors.

Massari et al. examined 151 GCs for which they collected complete kinematic information. They concluded that 62 of these clusters were most likely
formed in the Galaxy (in--situ), while the remaining clusters (89 clusters) were most likely formed ex--situ and then accreted. Basically, accreted clusters are associated with four known merger events: Gaia--Enceladus -- 26 GCs (+6 candidates), the Sagittarius dwarf galaxy -- eight GCs, the Helmi stream (H99) -- 10 GCs, and the Sequoia galaxy -- seven GCs. The remaining 36 clusters are classified as ``Low energy'' ones (25 GCs) or ``High energy'' ones (11 GCs). Association of the clusters with any group is uncertain, due to the partial overlap of the debris of different progenitor galaxies.

Myeong et al. considered 34 GCs, which accreted onto the Galaxy. To verify this, the authors used kinematic data of GAIA \citep{2018A&A...616A...2L}  in combination with photometry from DECaPS (DECam Plane Survey \citep{2018ApJS..234...39S}). In opinion of Mueong et al., 6 GCs belong to the Sagittarius dwarf spheroidal galaxy, 7 -- to Sequoia galaxy, 21 -- to the Gaia Sausage. Summing up the results of three above--mentioned papers, we obtain the main list of tidal streams from which accreted a considerable fraction of GCs:

(1) Sagittarius dwarf spheroidal galaxy ($Sgr$ $dSph$) with the nucleus $NGC$~6715 ($M54$).

(2) Sequoia galaxy with the nucleus $NGC$~5139 (Omega Centauri ($\omega$ Cen)).

(3) Helmi stream ($H99$).

(4) Gaia--Enceladus with the nucleus $NGC$~1851. Other possible variations of the name of this stream -- Gaia Sausage or Canis Major ($CMa$).

(5) Low--energy progenitor Coala, to which Kraken may be equivalent, and also a low--energy group ($E<-1.86\times10^5$ km$^2$ s$^{-2}$).

(6) High energy group ($E>-1.5\times 10^5$ km$^2$ s$^{-2}$).

\subsection{Anisotropy of the distribution of globular clusters}\label{sec:2.1}
The number of GCs belonging to different streams, according to the classification of the authors considered in this paper, is as follows: according to Forbes -- 87 GCs, according to Massari -- 89 (these clusters are located at the distance from 1.42 to 144.77 kpc from the center of the Galaxy) and according to Myeong -- 34 GCs (at the distance from 2.42 to 71.36 kpc).

In order to understand whether there is any difference in the distributions of GCs belonging to the streams and the objects formed in the Milky Way, we decided to check the anisotropy of the distribution of these GCs using the gyration tensor, as in \cite{2018MNRAS.481..918A}. The tensor is constructed as follows:

\begin{equation}
\label{form:1} 
 S_{ij}=\frac{1}{N}\sum_{k=1}^Nx_i^kx_j^k,
 \end{equation} 
 
where $S$ -- gyration tensor, $N$ -- the number of objects,  $x_i^k$ -- the distance of $k$ th object to the Galactic center along coordinate axis $i$. Standard mathematical operations for determination of the eigenvalues and eigenvectors of a tensor allow us to characterize the anisotropy of the distribution. The eigenvalues a, b, and c, for convenience, are sorted in ascending order, so that $a>b>c$ . The degree of anisotropy is characterized by the ratios of the eigenvalues $c/a$ and $b/a$, which, in the case of an isotropic distribution, approach 1. The eigenvectors of the inertia tensor determine the orientation of the anisotropic distribution in space.

To check the statistical significance of the found parameters of the GCs system, we generated 10 000 random samples with the same radial distribution and number of objects as in the data for observed objects, and measured the median value and the root-mean-square value of the ratio of the eigenvalues of the tensors. Anisotropy is statistically significant if the ratio of the eigenvalues of the tensor for the real catalogs differs from the median of the random samples by more than $3\sigma$. Random samples are constructed by fixing the distances ($R$) from the real sample and assigning random angular coordinates to the GCs.

\begin{figure*}
% \setcaptionmargin{5mm} \onelinecaptionstrue \captionstyle{normal}    
 \includegraphics[width=0.325\textwidth]{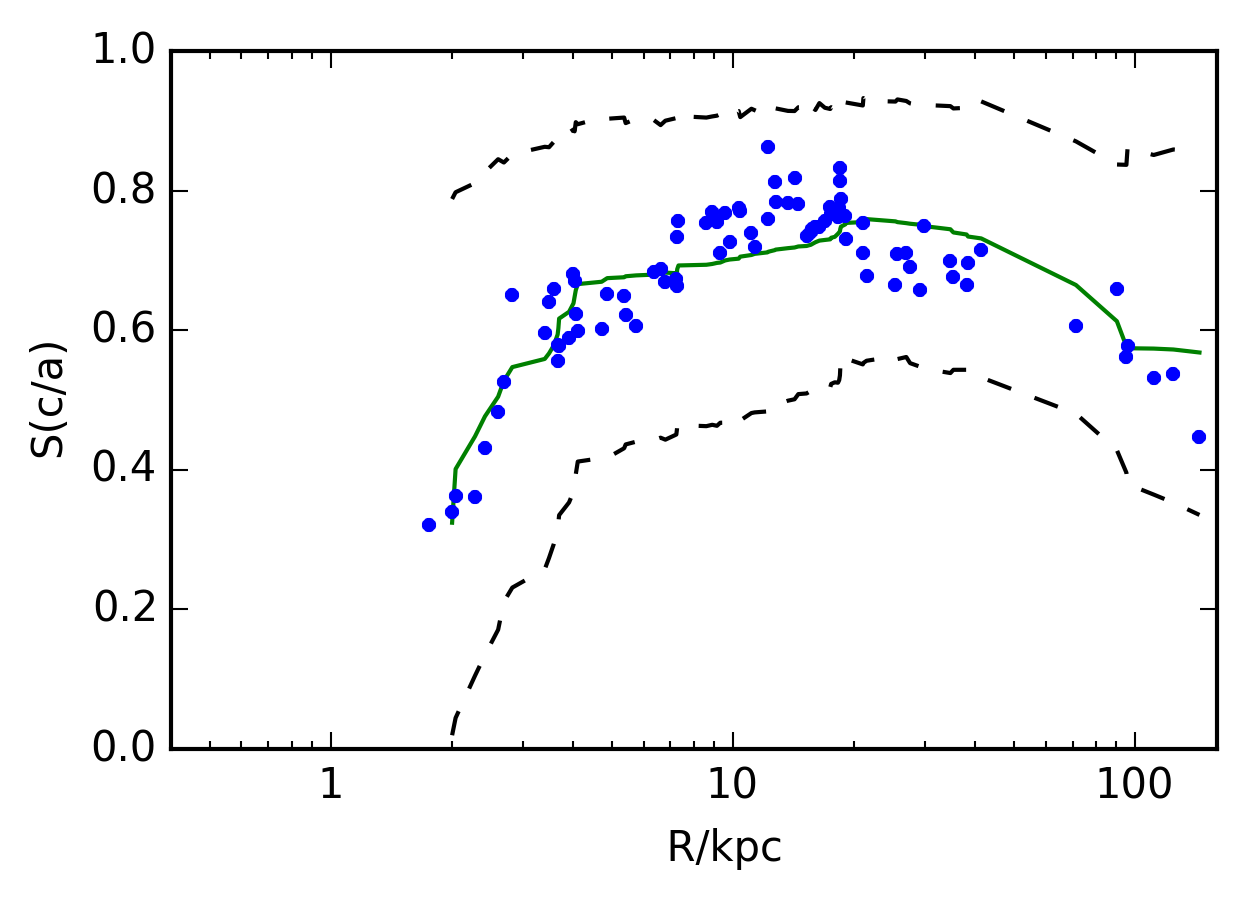}
 \includegraphics[width=0.325\textwidth]{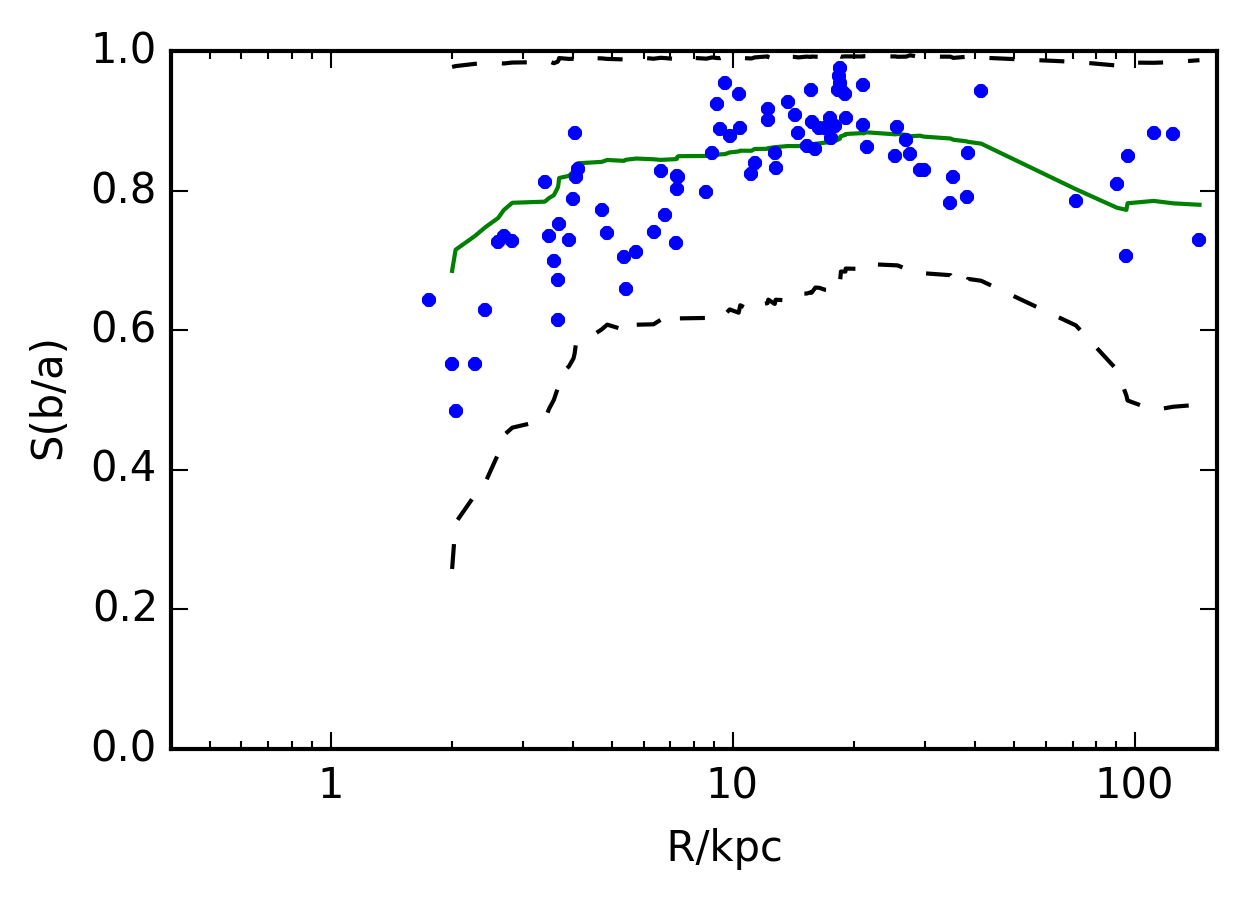}
 \includegraphics[width=0.325\textwidth]{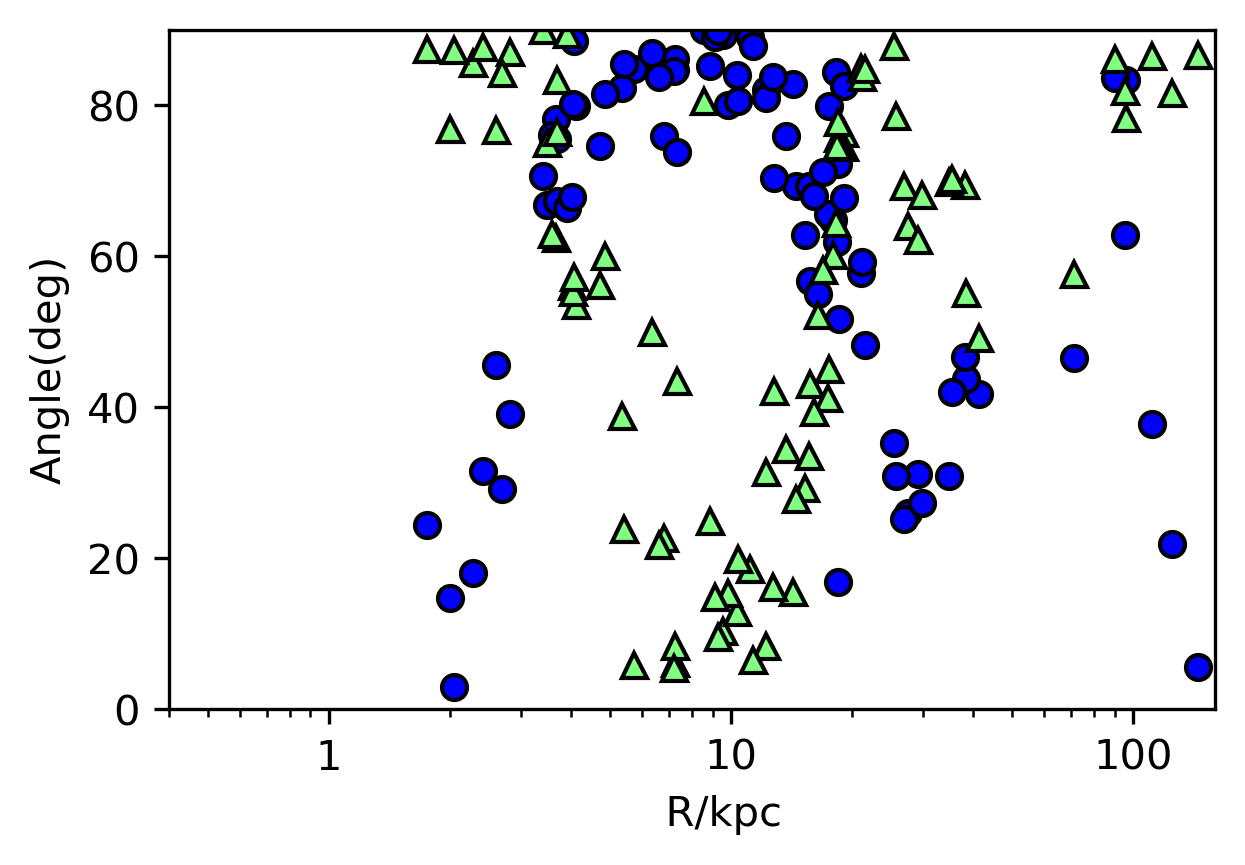} 
 
 \includegraphics[width=0.325\textwidth]{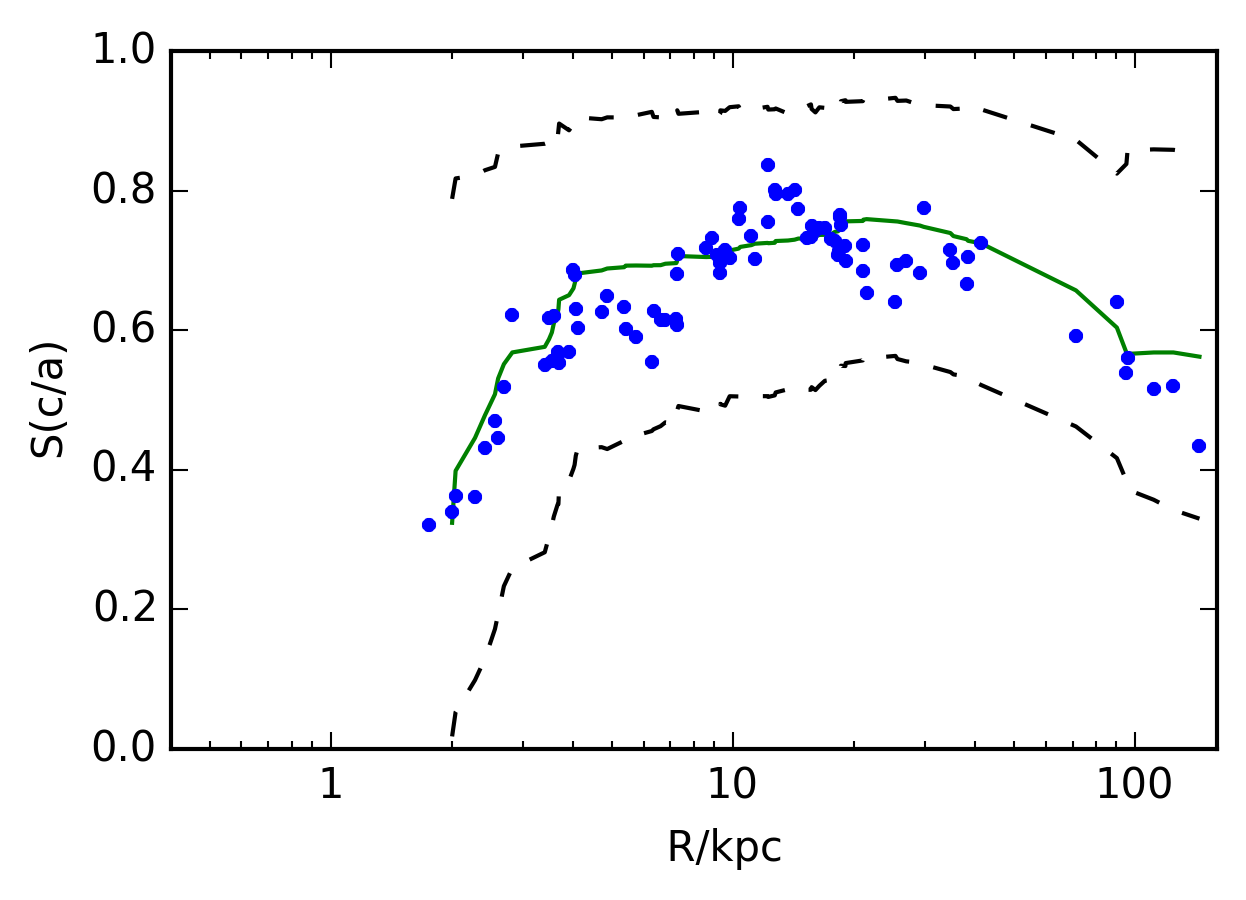}
 \includegraphics[width=0.325\textwidth]{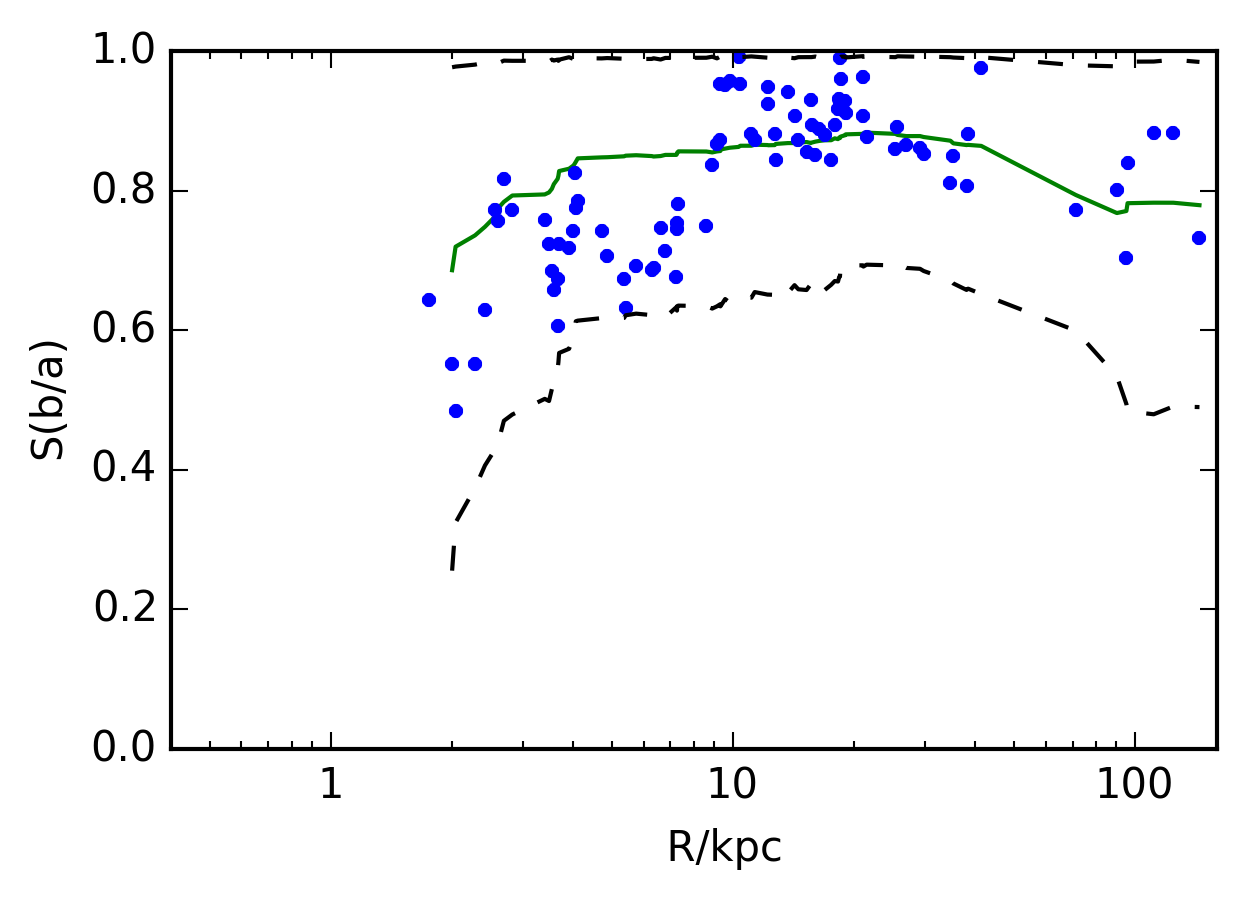}
 \includegraphics[width=0.325\textwidth]{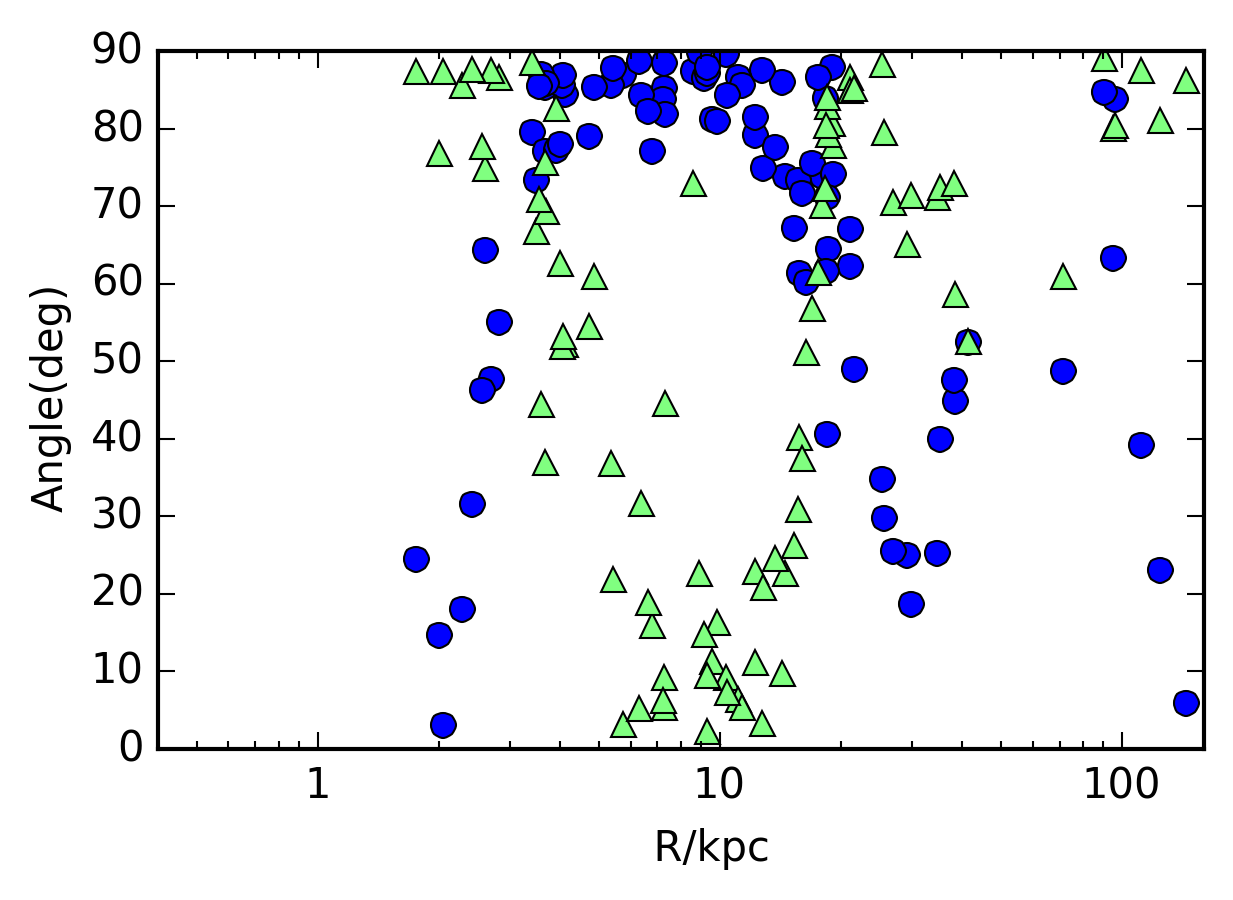}
 
 \includegraphics[width=0.325\textwidth]{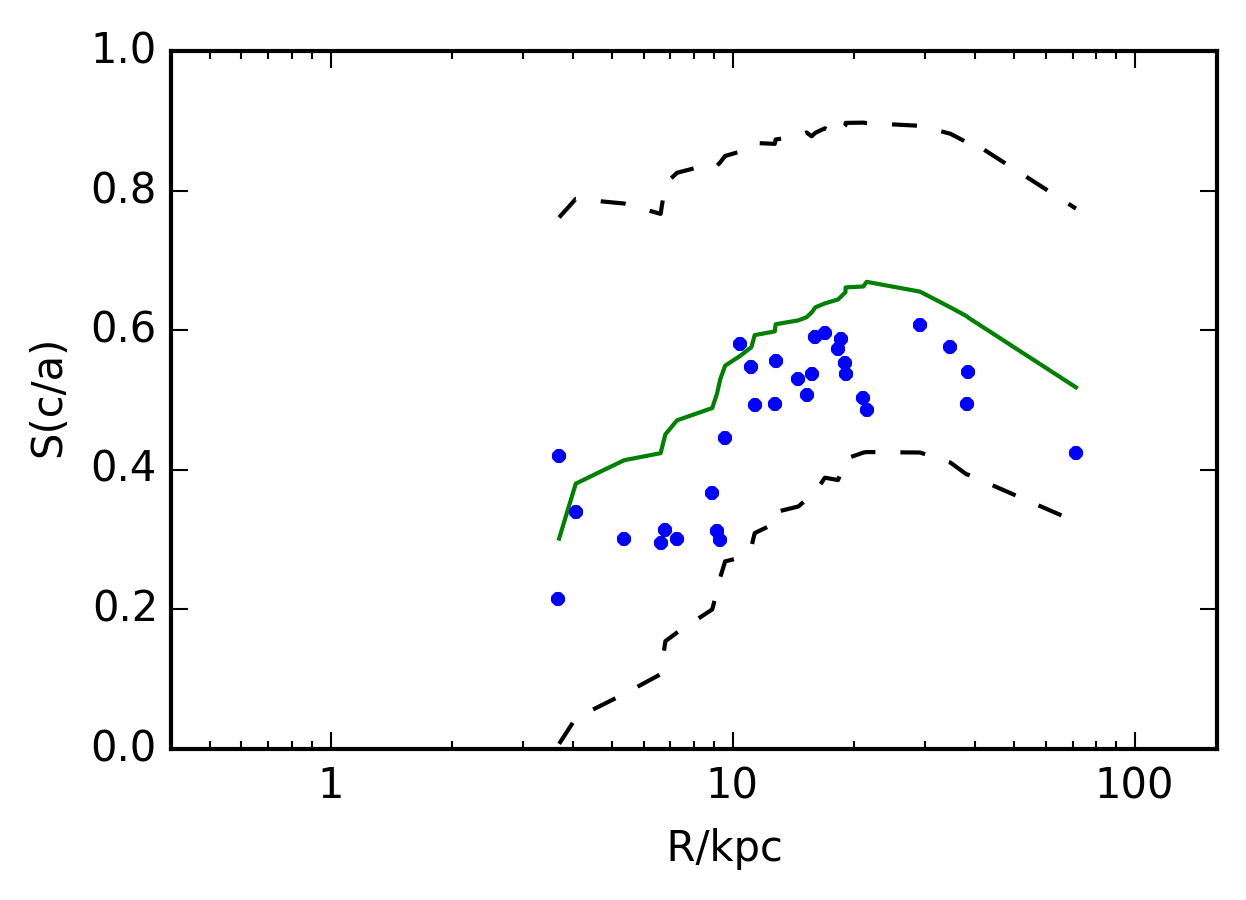}
 \includegraphics[width=0.325\textwidth]{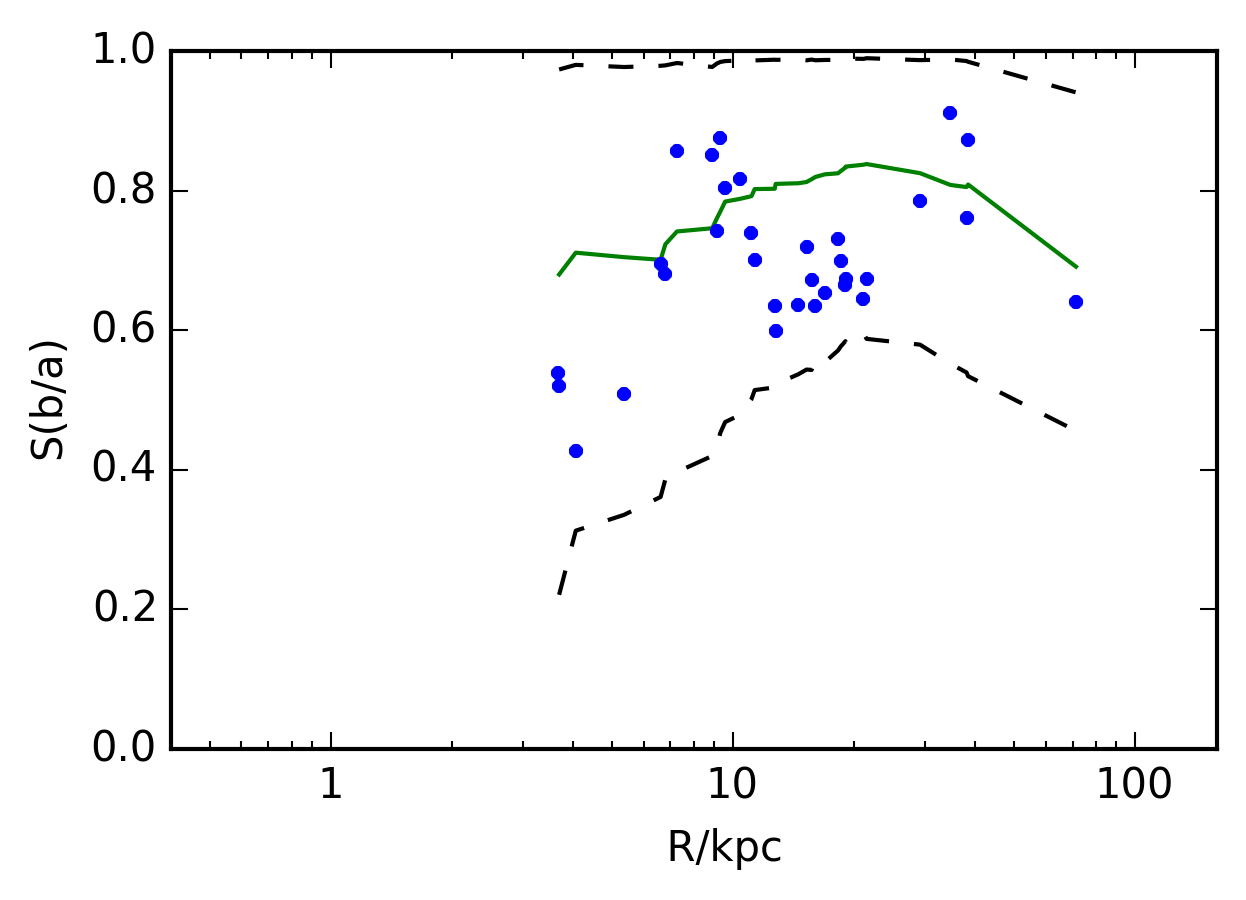}
 \includegraphics[width=0.325\textwidth]{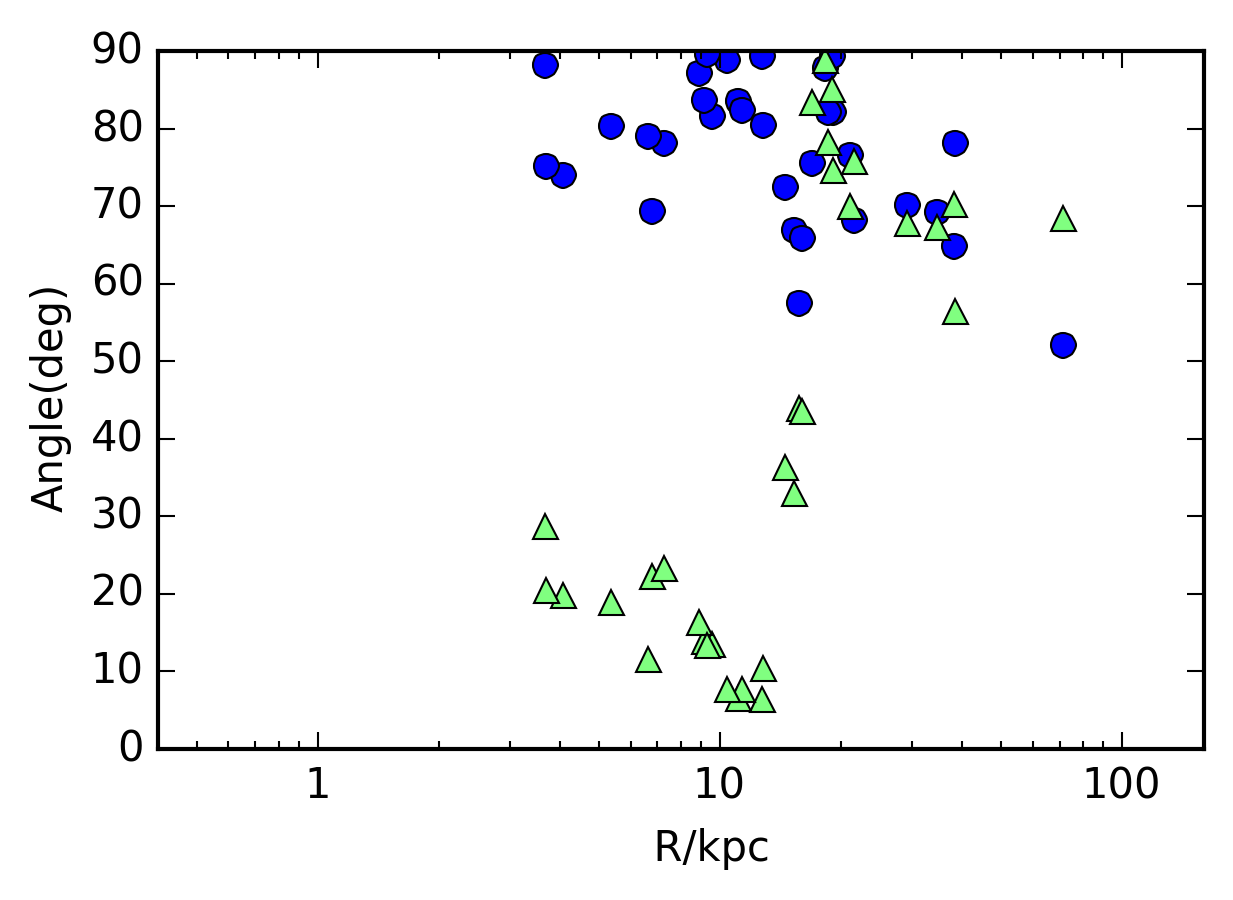}
 \caption{Galactic clusters anisotropy, quantified by the gyration tensor for cluster samples Forbes (top), Massari (center), and
Myeong (bottom). The left and middle columns show the ratios $c/a$ and $b/a$ as a function of the galactocentric cluster distance,
respectively. Each blue dot represents the ratio of the tensor eigenvalues calculated for all clusters at the distance smaller than R
from the Galactic center. Solid green line represents the median eigenvalues ratios for 10 000 random samples. Dashed lines indicate the deviations $\pm3\sigma$ of such random distributions. The right column shows the angles, measured in degrees, between the
Milky Way’s Galactic pole and the major (blue dots) and minor (green triangles) axis of the gyration tensor. Green triangles close
to $90^\circ$, indicate the polar plane. }
 \label{fig:1}
\end{figure*}

In Fig.~\ref{fig:1}, we show the results of measuring the anisotropy for GCs using the gyration tensor. The panels show the ratios $c/a$ and $b/a$ as functions of $R$ , calculated for all GCs with a distance $<R$ . The distributions of real objects are shown by dots, solid line represents the median result for 10 000 random samples, and dashed lines represent the median $\pm3\sigma$. The ``angle'' in these panels is measured between the normal to the Galactic plane and the minor (green triangles) or the major (blue dots) axis of distribution.

% Fig 1

From the measurements of the anisotropy using the gyration tensor for all samples of GCs in the streams from the three above-mentioned papers, it follows that the distribution of GCs in the streams is isotropic. Thus, none of the samples exhibit statistically significant anisotropy. In \citep{2018MNRAS.481..918A} [p. 7, Fig. 7], for the entire sample of GCs at the distance from 2 to 10 kpc, statistically significant anisotropy is observed, which the authors associated with GCs belonging to the disk of the Galaxy, that is, formed in--situ. In this paper, we studied spatial distributions of GCs, which, according to a number of authors, belong to the tidal streams, that is, were formed ex--situ. As seen in Fig.~\ref{fig:1}, for all samples, the spatial distribution of GCs belonging to the tidal streams is isotropic. This is consistent with the conclusion of Arakelyan et al. \citep{2018MNRAS.481..918A} that the statistically significant anisotropy for the entire GCs sample is due to the clusters that were most likely formed in the Galaxy or have been interacting with the Galaxy disk for a very long time. It is also important that the clusters that belong to the tidal streams do not exhibit significant structure, which we might expect, first, because clustering occurs mainly along the distinguished directions associated with the walls and filaments of a large--scale structure, and, second, because anisotropic distribution is observed for satellite galaxies.

Nevertheless, it is seen in Fig.~\ref{fig:1} that for all three samples (Forbes, Massari, Myeong) for GCs that belong to the streams, the major axis of the gyration tensor is in the disk, at the distances from about 3 to 10---20~kpc. It seems unlikely that such a situation can arise for a random isotropic GCs distribution. The distribution of the directions of the axes of the tensor as in Fig.~\ref{fig:1}, one can expect if a part of the GCs in each of the samples belongs to the disk. We demonstrate this below using random catalogs.

To check the probability of entering of the GCs from the disk into the GCs sample from the tidal streams, we generate random catalogs containing the same number of GCs as the real samples. Moreover, we take the galactocentric GC distances from the real samples. Angular coordinates are assigned randomly. To simulate a situation in which some of the clusters belong to the disk, for n clusters the height above the disk is set to zero (Cartesian coordinate $z$).

Using such models, we calculated the conditional probability of obtaining a distribution similar to the right-hand column in Fig.~\ref{fig:1}, i.e., when the major axes of the gyration tensor in the distance range from 3.5 to 20~kpc are located at an angle of more than $70^\circ$ to the direction to the Galactic pole, provided that $n$ clusters belong to the disk. If the distribution is isotropic, i.e. $n=0$, this probability is equal to 4.5, 0.6, and 1.1\% for the samples of Forbes, Massari and Myeong, respectively.

For this probability to exceed, for example, 10\%, the disk must contain $n=6$, 16, and 8 GCs for the Forbes, Massari, and Myeong samples, respectively. From this, we can conclude that a part of the GCs, formed to the opinion of these authors outside our Galaxy, actually belongs to its disk. It should be noted that in \citep{2020ARep...64..805M}, based on the analysis of the abundance of alpha--elements, it was shown that the group of Low energy clusters from the Massari work was most likely formed in--situ, which also indicates the inaccuracy of the in--situ/ex--situ separation in the Massari sample. In order to verify further the origin of the GCs, we use the ``age--metallicity'' diagram.

\section{TWO BRANCHES OF THE GSs IN THE ``AGE--METALLICITY'' RELATION}\label{sec:3}
The literature discusses the fact that the population of the GCs of the Milky Way exhibits bimodality of colors: there are blue and red clusters \citep{1978ApJ...225..357S,1993MNRAS.264..611Z,1993AJ....105.1762O,1995ApJ...454L..73W,1996MNRAS.280..971E,1999AJ....118.1526G,2001ApJ...556..801L,2001AJ....121.2974L,2006ApJ...639...95P,2006AJ....132.1593S,2006AJ....132.2333S,2013ApJ...762...39T,2017MNRAS.465.3622R}. This is due to the bimodality of metallicity \citep{1997AJ....113.1652F,1997AJ....113..887F,1998ApJ...501..554C,2001ApJ...563L.143F,2005A&A...439..997P,2005AJ....130.1315S,2005AJ....129.2643B,2006ARA&A..44..193B,2006MNRAS.366.1253P,2006AJ....132.2333S,2006ApJ...639...95P,2011MNRAS.413.2943F}. Blue clusters are found mostly in the halo of the Galaxy. These clusters probably previously belonged to the satellite galaxies. At the same time, red clusters are spatially concentrated towards the Galactic center and rotate with it. Blue clusters are old and metal-poor, while red clusters are younger and metal--rich. The ratio [$Fe/H$] peaks for blue and red clusters in the Milky Way are approximately –1.5 and –0.5, respectively. Such a bimodality assumes two mechanisms of GCs formation. The authors of \citep{1998ApJ...501..554C,2013ApJ...762...39T,2017MNRAS.465.3622R} argue that red clusters are formed in--situ, while blue ones were accreted either as a result of the merger of satellite galaxies with the Galaxy or as a result of tidal capture of the clusters themselves.

\begin{figure*}
% \setcaptionmargin{5mm} \onelinecaptionstrue \captionstyle{normal}    
 \includegraphics[width=0.325\textwidth]{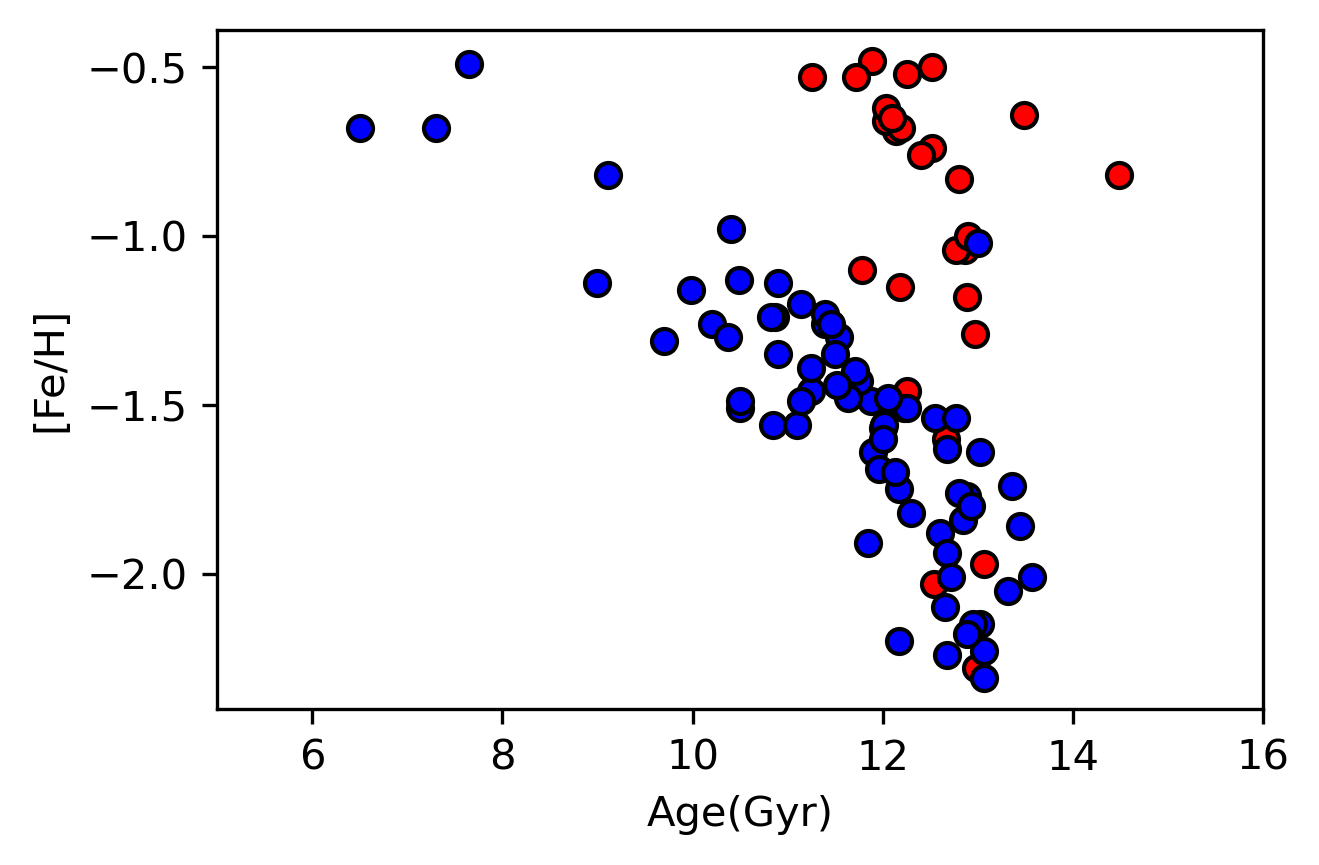}
 \includegraphics[width=0.325\textwidth]{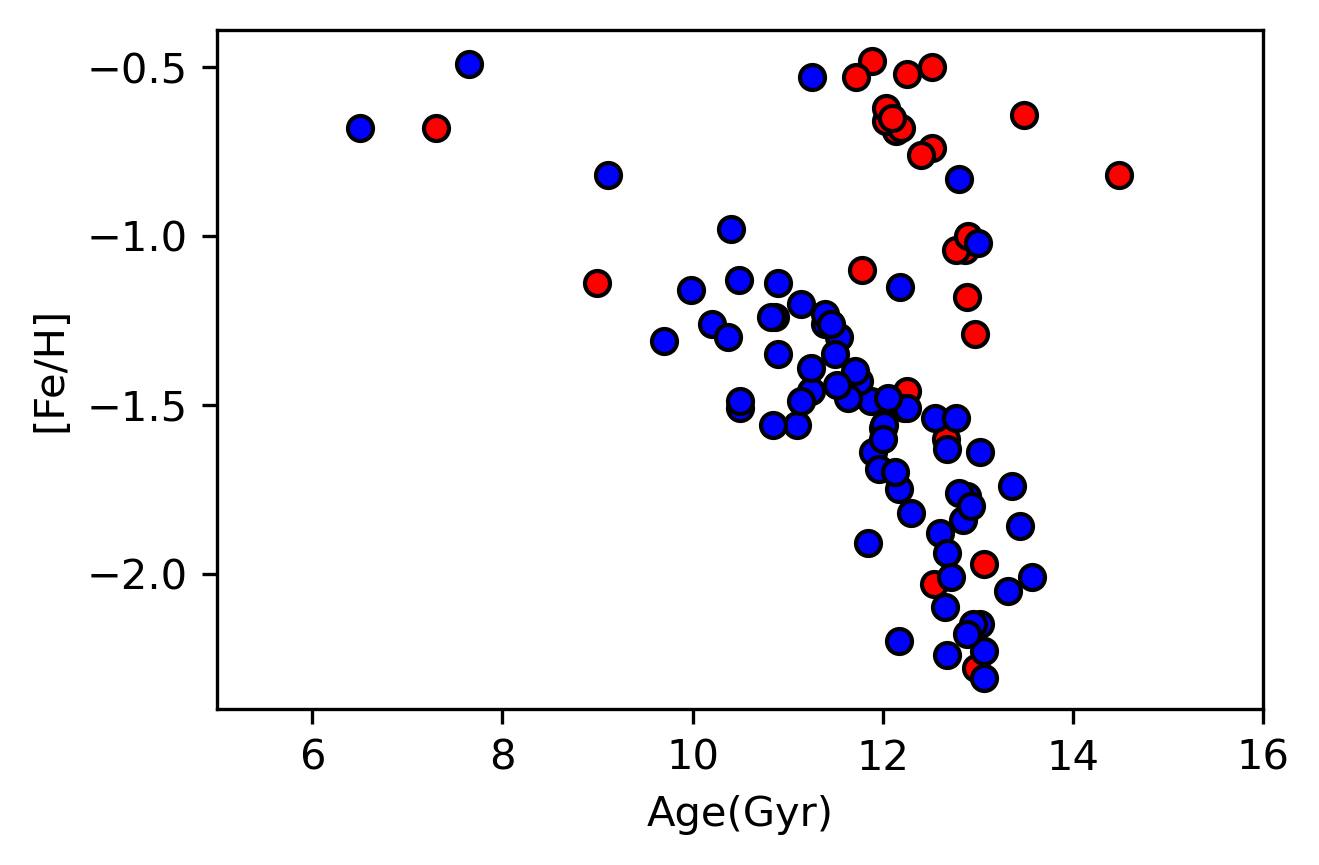}
 \includegraphics[width=0.325\textwidth]{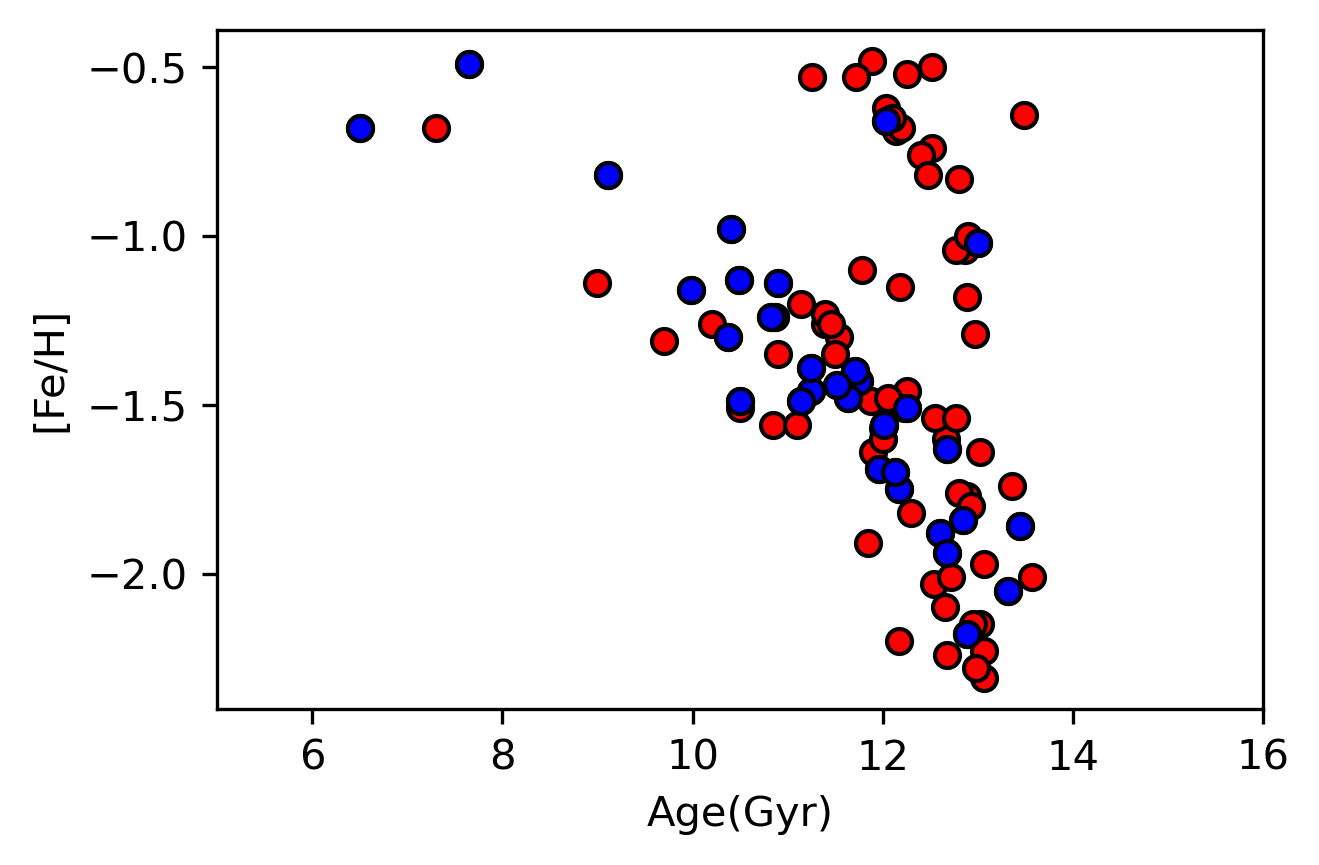} 
 \caption{Dependence of the age of GCs on metallicity for cluster samples Forbes, Massari, and Myeong, the left, middle, and right panels, respectively. Blue dots represent ex--situ clusters and the red ones show in--situ clusters.}
 \label{fig:2}
\end{figure*}

To understand the difference between in--situ clusters and ex--situ clusters, we plotted respective samples from Massari, Forbes, and Myeong in the ``age--metallicity'' diagram. The results are shown in Fig.~\ref{fig:2}.

% Fig 2

The  ``age--metallicity''  dependence  clearly  shows that GCs have two branches. The low-metals branch contains mainly clusters that belong to different tidal streams  formed  by  the  partial  destruction  of  satellite galaxies.  The  clusters  in  this  sequence  show  a  wide variation in age and metallicity, but there are no clusters less than 6 Gyr old. The clusters of a more metals-rich  branch,  formed  in--situ,  also  have  a  scatter  in metallicity, but all clusters are more than 11 Gyr old.

It  is  worth  noting  that  the  in--situ  clusters  were formed not in the Galaxy as we know it, but in its progenitor. In the hierarchical model of the formation of galaxies, the mass of a galaxy is accumulated gradually due to mergers, and galaxies as a whole do not have a clearly  defined  moment  of  formation.  Therefore,  for the  objects  formed  long  ago,  it  is  difficult  to  distinguish  between  the  concepts  of  in--situ  and  ex--situ. However, specifically for our Galaxy, it is believed that it did not experience mergers with the objects of comparable  mass  since  $z=2$  or  less  than  10.5  Gyr  ago \citep{2007ApJ...662..322H}. By that time, it had gained only 1/5 of its current total mass (including the dark halo). Six Gyr ago (the age of the youngest GCs), its mass was about 60\% ofthe current one \citep{2020MNRAS.491.1531C}.

\section{THE ROLE OF THE LOCAL SUPERCLUSTER}\label{sec:4}
In  the  hierarchical  model  of  galaxies  formation, accretion  of  the  matter  is  controlled  by  large--scalef lows, which are also responsible for the formation of a  cellular  structure--``pancakes''  and  filaments.  In accordance with Zeldovich theory \citep{1970A&A.....5...84Z}, a ``pancake'' is formed from a uniformly filled volume if compression occurs  in  one  of  the  three  mutually  perpendicular directions,  and  expansion --- in  two  other  directions. Thus,  the  large--scale  structure  is  associated  with anisotropic motions of matter, and this anisotropy can also affect the distribution of the matter in the galaxies.  Our  Galaxy,  together  with  the  Local  Group,  is located within the Local Supercluster (LSC) \citep{1953AJ.....58...30D,1956VA......2.1584D,1975ApJ...202..610D,1975ApJ...202..616D,1976RC2...C......0D,1991rc3..book.....D}, a well--visible pancake--like structure with dimensions of tens of Mpc.

\begin{figure*}
% \setcaptionmargin{5mm} \onelinecaptionstrue \captionstyle{normal} 
\begin{center}
 \includegraphics[width=0.325\textwidth]{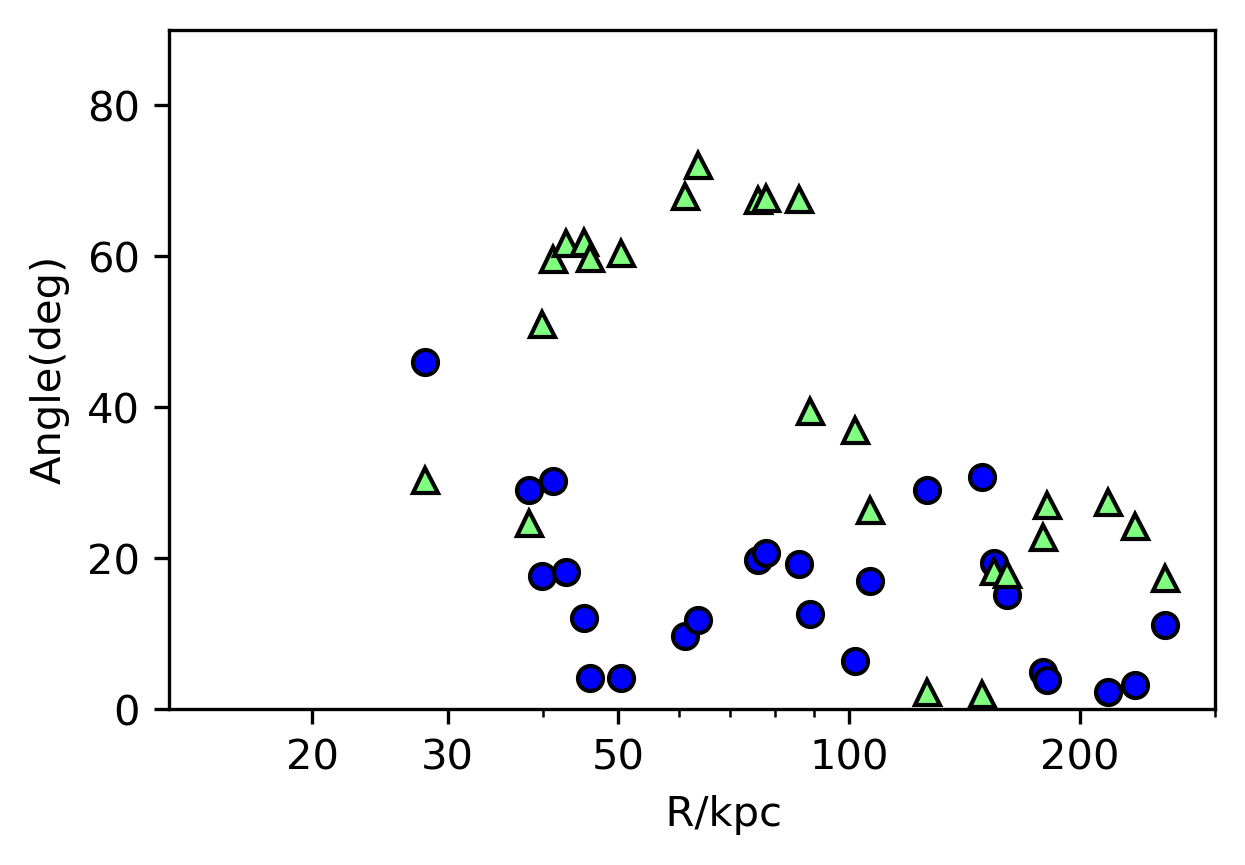}   
 \includegraphics[width=0.325\textwidth]{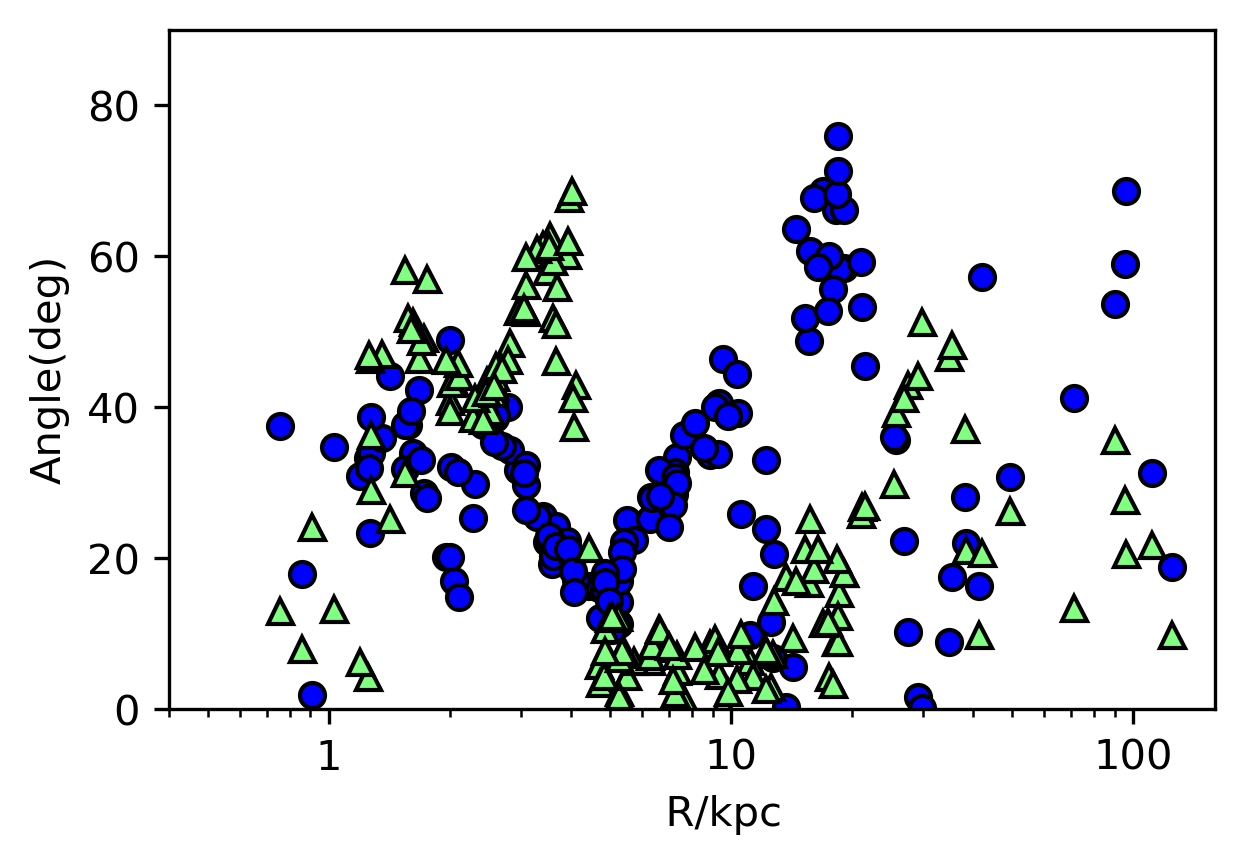}  
 \end{center}
 \includegraphics[width=0.325\textwidth]{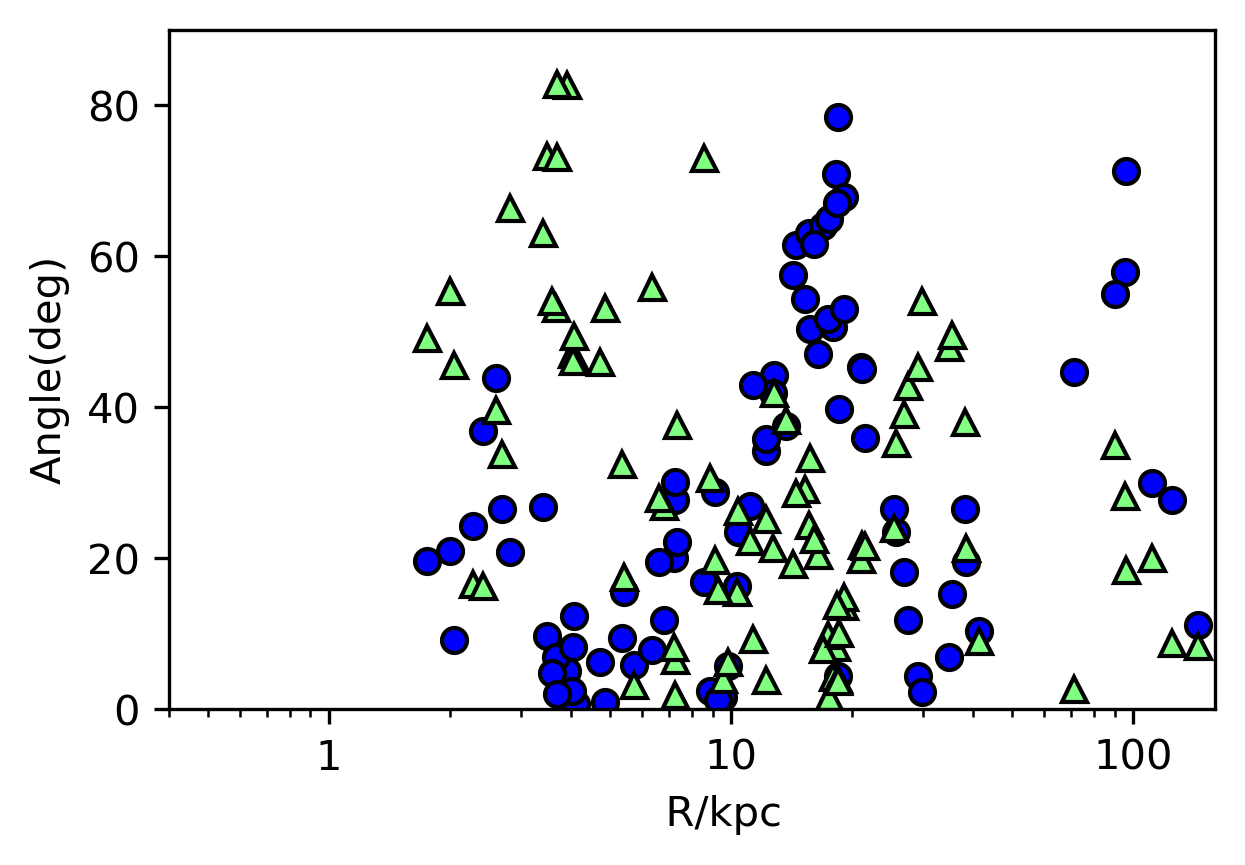}
 \includegraphics[width=0.325\textwidth]{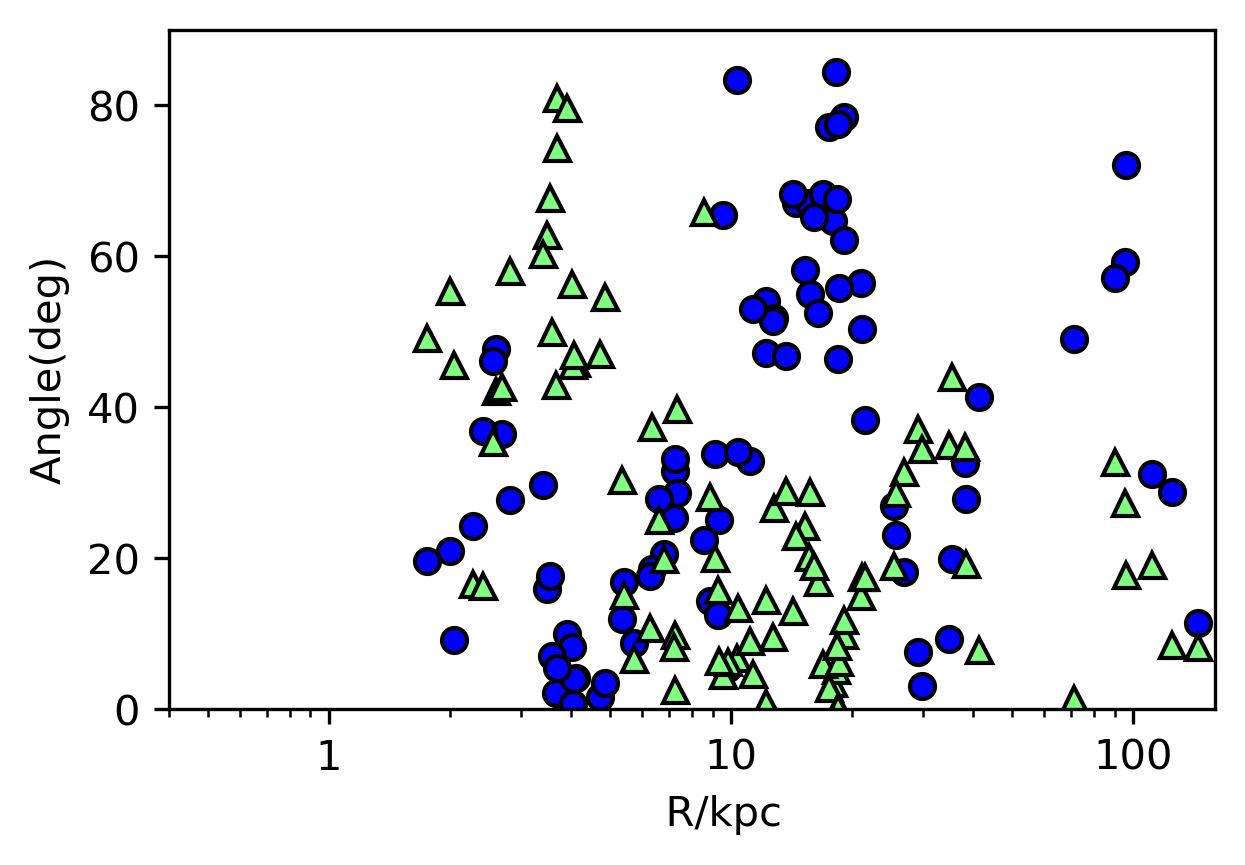}
 \includegraphics[width=0.325\textwidth]{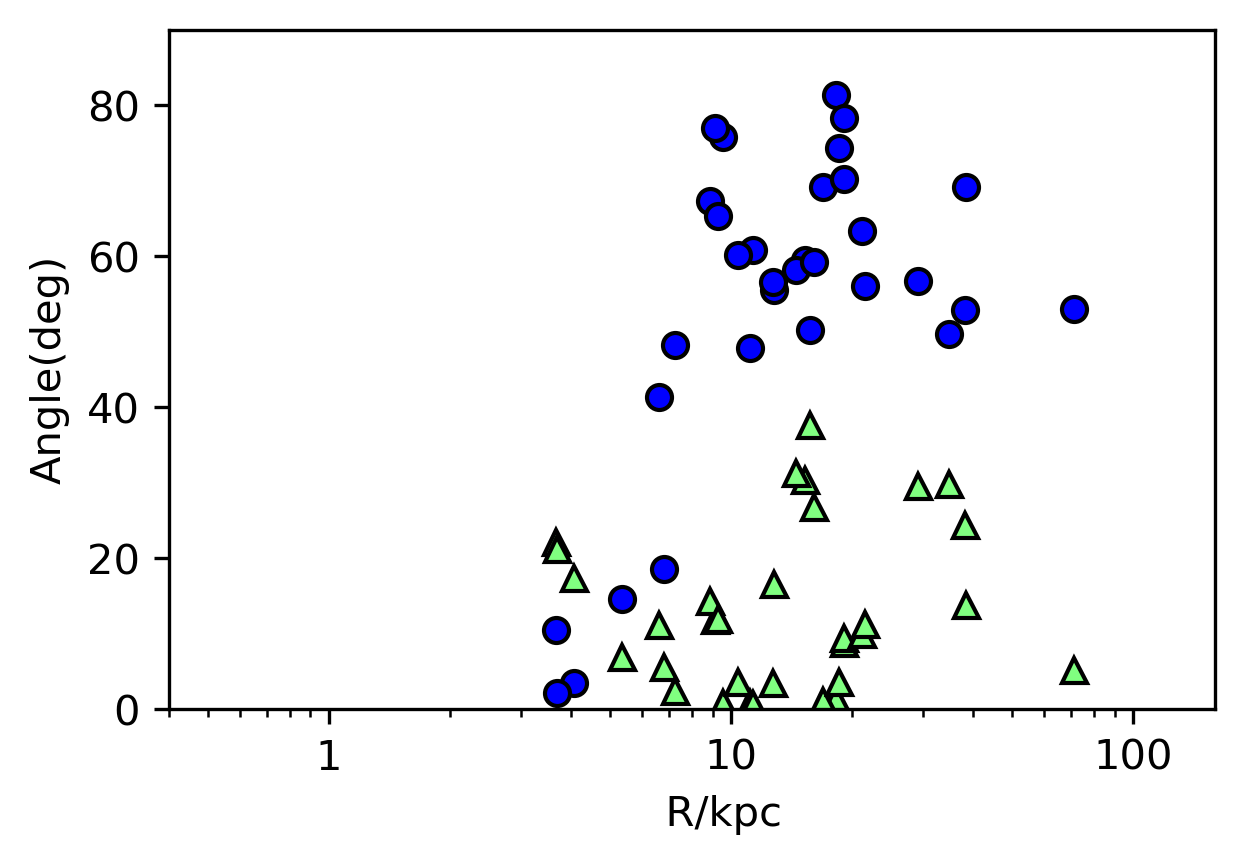}
 \caption{The ``Angle'' as a function of galactocentric distance for satellite galaxies for entire GCs sample (top row, left to right) and for cluster samples Forbes, Massari, and Myeong (bottom row, left to right, respectively). The ``Angle'' is measured between the LSC plane and the small (green triangles) or large (blue dots) axes of the distribution of GCs.  }
 \label{fig:3}
\end{figure*}

We  have  tested  the  influence  of  the  Local  Supercluster  on  the  spatial  distribution  of  GCs,  as  well  as dwarf satellite galaxies of the Milky Way. The satellite galaxies  were  a  priori  accreted  onto  our  Galaxy  from the  outside.  At  the  same  time,  they  form  a  clear--cut flat  structure  \citep{2005A&A...431..517K,2008ApJ...680..287M,2018MNRAS.481..918A}.  Therefore,  we  did  not  limit ourselves  to  analyzing  the  GCs  distribution,  but  also considered satellite galaxies. For this, the angle distributions  between  the  axes  of  the  gyration tensor   (\ref{form:1}) and the plane of the Local Supercluster were obtained for  dwarf  satellite  galaxies  (27  satellites  \citep{2012AJ....144....4M})  and  for: (i) for the entire GCs sample (157 GCs \citep[][2010 edition]{1996AJ....112.1487H}, \citep{2013ApJ...772...82H}); (ii) for the  GCs  from  the  Forbes  list;  (iii)  for  the  GCs  from the Massari list, and (iv) for the GCs from the Myeong list. In Fig.~\ref{fig:3}, the ``Angle'' is presented as a function of galactocentric  distance  for  the  GCs  and  satellites  of the  Galaxy.  The  ``Angle''  is  measured  between  the plane of the Local Supercluster and the minor (green triangles) or major (blue dots) axes of the distribution of GCs.

% Fig 3

Fig.~\ref{fig:3}  shows  that  for  satellite  galaxies  (top  row, left) at the largest distances, the major and minor axes are  located  in  the  supergalactic  plane.  At  the  same time, the minor axis is located in the disk of the Milky Way  and  the  major  one  is  perpendicular  to  the  disk. This  means  that  the  plane  of  satellite  galaxies  is  perpendicular  to  both  the  disk  of  the  Galaxy  and  the supergalactic plane.

We can say the following regarding all 157 GCs (top row, on the right). At small distances, up to 4~kpc, we are not interested in the result, since these are GCs in the central part of the Galaxy. From 4 to 20~kpc, the minor axis of the system is perpendicular to the disk of the Galaxy and is in the supergalactic plane. The major axis is in the disk of the Galaxy and is perpendicular to the supergalactic plane at a distance of about 20~kpc. Thus, in the range from 4 to 20~kpc, the orientation of the GC system corresponds to the disk of the Galaxy, the influence of the Supercluster does not manifest itself. The same can be said for the clusters from streams, but only with a caveat that they have a larger noise. The minor axis shows a large scatter; this may be due to the fact that there are fewer clusters belonging to the disk of the Galaxy in the sample of objects from the streams.

At the distance of about 100~kpc, the picture for GCs resembles that of satellite galaxies for all samples, i.e., the GC system is oriented perpendicular both to the disk and the Supercluster. It is worth noting that only six clusters are observed at such distances, which is not enough for reliable conclusions.

At a distance of about 30~kpc in all samples except Myeong, the major axis is in the supergalactic plane, while the minor one makes with the large axis an angle of about $60^\circ$ for all GCs and for Forbes sample; for the Massari sample, the angle is within $45^\circ$. At a distance of 25--40~ kpc, there are only 10 GCs, of which 10 and 9 GCs belong to the streams of the Forbes and Massari samples, respectively. Thus, in the Forbes and Massari samples at these distances, there may be signs of the influence of the Supercluster on the orientation of the system of accreted GCs, but the reliability of this conclusion is low.

\section{DISCUSSION AND CONCLUSIONS}\label{sec:5}
In this paper, we studied the GC system that was formed outside the Galactic disk. To do this, we took from the literature the samples of GCs that were formed in different tidal streams. We chose the works of Forbes, Massari, and Myeong, since their lists of GCs belonging to the different streams are the most complete and are based on the latest data from the GAIA observatory. Having studied a number of works, including those mentioned above, we obtained the main list of tidal streams in which GCs belonged and later were accreted: Sagittarius dwarf spheroidal galaxy (Sgr dSph), Sequoia Galaxy (Sequoia), Helmi Stream (H99), Gaia--Enceladus (possibly Gaia Sausage or CMa), the low energy group (possibly Koala or Kraken), and the High energy group.

It is believed that the accretion onto the Galaxy was anisotropic, which is manifested, for example, as a disk--like structure of satellite galaxies. We measured the anisotropy of the distribution of GCs that belonged to the streams using the gyration tensor. The measurement result showed that no statistically significant anisotropy is observed for accreted GCs. Having obtained this result, we can state that the anisotropic structure that is observed for the complete sample of GCs (see \citep{2018MNRAS.481..918A}, p. 7, Fig. 7) is due to the presence of many GCs in the Galactic disk, and is associated with the clusters formed in--situ.

However, in Fig.~\ref{fig:1} for the three samples of the accreted GCs, the major axis of the gyration tensor at a distance from 3 to 20~kpc is in the disk. This may be due to the fact that the samples contain a significant number of GCs that have formed in the disk of the Galaxy. To estimate their number, the distribution of GCs with random angular coordinates was modeled and it was shown that the probability of a random realization of such a distribution, in which there are no GCs belonging to the disk, is 4.5, 0.6, and 1.1\% for the Forbes, Massari, and Myeong samples, respectively. This conclusion is consistent with the conclusion of Marsakov et al. \citep{2020ARep...64..805M}, who had shown that some of the
clusters from the Massari catalog claimed to be ex--situ are in fact genetically related to our Galaxy.

We also checked how the clusters formed in--situ and ex--situ behave respective to the AMR (Fig.~\ref{fig:2}). Two branches can be easily distinguished; the low--metals branch contains mainly clusters belonging to different streams, and they have a large spread in the age and metallicity. At the same time, the clusters in the more metallic branch, which most likely formed in the Galaxy, have a scatter in metallicity, but their age is over 11~Gyr.

To check the likely influence of the Local Supercluster on the distribution of satellite galaxies and GCs of the Milky Way, we presented the Figures, which show the angle between the LSC plane and the axes of distribution of GCs systems or satellite galaxies, as a function of the galactocentric distance. Fig.~\ref{fig:3} (top row, left) shows that the plane of the satellite galaxies is both perpendicular to the disk of the Galaxy and to the supergalactic plane. For GCs at the distances of up to 20~kpc, only the influence of the Galactic disk is traced; at the distances of about 30~kpc, the orientation of the GCs system may coincide with the supergalactic plane, and at larger distances (more than 100~kpc), the orientation resembles that for satellite galaxies.

% *** Астрофизические исследования % русск. версия журнала САО до 1993 г.
\newcommand{\air}{Астрофиз. исслед. (Известия Спец. астрофиз. обс.) }
% *** Astrophysical Bulletin % англ. версия журнала САО с 2007 г.
\newcommand{\ab}{Astrophysical Bulletin }
% *** Астрофизический бюллетень % русск. версия журнала САО с 2007 г.
\newcommand{\abr}{Астрофизический бюллетень }
% *** Astronomy and Astrophysics
\newcommand{\aaa}{Astron. and Astrophys. }
\newcommand{\aap}{Astron. and Astrophys. }
% *** Astronomy and Astrophys. Supplement Series
\newcommand{\aas}{Astron. and Astrophys. Suppl. }
\newcommand{\aaps}{Astron. and Astrophys. Suppl. }
% *** Astronomy and Astrophysics Review
\newcommand{\aar}{Astron. Astrophys. Rev. }
% *** Astronomical Journal
\newcommand{\aj}{Astron.~J. }
% *** Astrophysical Journal
%\newcommand{\apj}{Astrophys.~J. }
% *** Astrophysical Journal Supplement Series
\newcommand{\apjs}{Astrophys.~J. Suppl. }
% *** Astrophysics and Space Science
\newcommand{\apss}{Astrophys. and Space Sci. }
% *** Annual Review of Astronomy and Astrophys.
\newcommand{\araa}{Annual Rev. Astron. Astrophys. }
% *** Astronomicekij Zhurnal
\newcommand{\azh}{Astron.~Zh. }
% *** Bulletin of the American Astron. Society
\newcommand{\baas}{Bull. Amer. Astron. Soc. }
% *** Bulletin of the Special Astrophysical Observatory % англ. версия до 2007 г.
\newcommand{\bsao}{Bull. Spec. Astrophys. Obs. }
% *** Бюллетень Спец. астрофизич. обсерватории % русск. версия до 2007 г.
\newcommand{\bsaor}{Бюлл. Спец. астрофиз. обсерв. }
% *** Inform. Bul. Var. Stars
\newcommand{\ibvs}{Inform. Bull. Var. Stars }
% *** Journal of Astronomy and Astrophysics
\newcommand{\jaa}{J.~Astron. Astrophys. }
% *** Monthly Notices of the Roy. Astron. Society
\newcommand{\mnras}{Monthly Notices Royal Astron. Soc. }
% *** Publ. of the Astron. Society of Australia
\newcommand{\pasa}{Publ. Astron. Soc. Australia }
% *** Publ. Astronom. Soc. Japan
\newcommand{\pasj}{Publ. Astron. Soc. Japan }
% *** Publ. of the Astron. Society of the Pacific
\newcommand{\pasp}{Publ. Astron. Soc. Pacific }
% *** Astronomy Reports (АЖ)
\newcommand{\arep}{Astronomy Reports }
% *** Astronomy Letters (ПАЖ)
\newcommand{\alet}{Astronomy Letters }
% *** Astronomische Nachrichten
\newcommand{\an}{Astronomische Nachrichten }
% *** Pis'ma v Astronomicekij Zhurnal
\newcommand{\pazh}{Pis'ma Astron. Zh. }
% *** Письма в АЖ
\newcommand{\pazhr}{Письма в АЖ }
% *** Астрон. ж.
\newcommand{\azhr}{Астрон.~ж. }
% *** Soviet Astronomy
\newcommand{\sovast}{Sov. Astron. }
% *** Scientific American
\newcommand{\sca}{Scientific American }
% *** Sky and Telescope
\newcommand{\skytel}{Sky Telesc. }
% *** Space Science Reviews
\newcommand{\spsrev}{Space Sci.~Rev. }
% Revista Mexicana de Astronomia y Astrofisica
%\newcommand{\rmxaa}{Revista Mexicana de Astronom\'{\i}a y Astrof\'{\i}sica}
\newcommand{\rmxaa}{Revista Mexicana Astronom. Astrof\'{\i}s. }
\newcommand{\apjl}{Astrophys.~J. Lett. }
%\newcommand{\nat}{Nature }
% *** Physical Review D
%\newcommand{\prd}{Phys. Rev.~D }
\newcommand{\memsai}{Memorie della Societ\`a Astronomica Italiana}

%\bibliography{gc.bib} 

\end{document}